\documentclass[12pt]{iopart}

\usepackage{iopams}

\usepackage{graphicx}
\usepackage{dcolumn}
\usepackage{bm}

\newcommand{\Tc}{$T_c$}

\begin{document}

\title{Superconductivity in $Ln$FePO ($Ln$ = La, Pr, and Nd) single crystals}

\author{R.\ E.\ Baumbach, J.\ J.\ Hamlin, L.\ Shu, D.\ A.\ Zocco,\\ N.\ M.\ Crisosto, M.\ B.\ Maple}

\address{Department of Physics and Institute for Pure and
Applied Physical Sciences, University of California, San Diego, La
Jolla, CA 92093}

\ead{mbmaple@ucsd.edu}

\begin{abstract}
Single crystals of the compounds LaFePO, PrFePO, and NdFePO have been prepared by means of a flux growth technique and studied by electrical resistivity, magnetic susceptibility and specific heat measurements.  We have found that PrFePO and NdFePO display superconductivity with values of the superconducting critical temperature \Tc\ of 3.2 K and 3.1 K, respectively.  The effect of annealing on the properties of LaFePO, PrFePO, and NdFePO is also reported.  The $Ln$FePO ($Ln$ = lanthanide) compounds are isostructural with the $Ln$FeAsO$_{1-x}$F$_x$ compounds that become superconducting with \Tc\ values as high as  55 K for $Ln$ = Sm.  A systematic comparison of the occurrence of superconductivity in the series $Ln$FePO and $Ln$FeAsO$_{1-x}$F$_x$ points to a possible difference in the origin of the superconductivity in these two series of compounds.
\end{abstract}

\maketitle

\section{Introduction}
Although the rare-earth transition metal phosphide oxides
$Ln$FePO, where $Ln$ is a lanthanide, were first synthesized in
1995 \cite{zimmer_1995_1}, it was not until 2006
\cite{kamihara_2006_1} that superconductivity near 5 K was
discovered in LaFePO.  The subsequent report of superconducting
critical temperature \Tc\ values near 26 K in the related,
fluorine-doped compound LaFeAsO$_{1-x}$F$_x$ \cite{kamihara_2008_1} resulted in a great
deal of excitement and many publications on what is now
recognized as a new class of Fe-based high temperature
superconductors.  Several different types of doping have been
found to induce superconductivity in these compounds including
substituting F for O~\cite{kamihara_2008_1,chen_2008_1,ren_2008_3,ren_2008_2,fang_2008_2,ren_2008_1,yang_2008_1}, Co for Fe \cite{sales_2008_2}, Sr for La
\cite{zhu_2008_1}, Th for Gd~\cite{ren_2008_5}, and also the introduction of oxygen vacancies
\cite{yang_2008_1,ren_2008_6}.  Materials with related structures which
include the same type of Fe-$Pn$ planes such as
Ba$_{1-x}$K$_x$Fe$_2$As$_2$ \cite{johrendt_2008_1} and LiFeAs
\cite{chu_2008_1} have also been found to exhibit similar physics
including relatively high temperature superconductivity.  Clearly,
there exists an enormous phase space of potential Fe-$Pn$
superconductors.  The vast majority of work has focused on the
arsenic-based versions of these compounds since, thus far, they
exhibit the highest \Tc\ values, with the record high \Tc\ near 55
K for SmFeAsO$_{1-x}$F$_x$ \cite{ren_2008_1} and
SmFeAsO$_{1-\delta}$ \cite{ren_2008_6}, and 56 K for
Gd$_{1-x}$Th$_x$FeAsO \cite{ren_2008_5}.  Given the toxic nature
of arsenic, it would be highly desirable to discover similarly
high \Tc\ values in less toxic P-, Sb-, or Bi-based versions of
these compounds.  However, despite the fact that replacing La with
the lanthanides Ce, Pr, Nd, Sm, or Gd in the As-based oxypnictides
leads to a near doubling of \Tc, there appears to be a striking
lack of data on the corresponding phosphorus based versions of
these compounds.  Although LaFePO has received substantial
attention, SmFePO was recently been reported to be superconducting
\cite{kamihara_2008_2}, and CeFePO remains non-superconducting
down to 400 mK \cite{bruning_2008_1}, the compounds $Ln$FePO with
$Ln$ = Pr, Nd, and Gd do not yet appear to have been scrutinized
for the occurrence of superconductivity.

The undoped parent compounds, $Ln$FeAsO, exhibit a spin
density wave (SDW) and structural instability near 150 K, remain
metallic to low temperatures, and only display superconductivity
when the SDW is suppressed towards zero temperature either through
doping or pressure~\cite{okada_2008_1}.  In contrast, the phosphorus-based analogues LaFePO
and SmFePO \cite{kamihara_2008_2} do not manifest a SDW transition
and appear to develop a superconducting state in the undoped form
at ambient pressure.  Three key questions are whether the mechanism
of superconductivity is the same in the P- and As- based compounds, whether the P-based versions of these compounds indeed exhibit
superconductivity in the undoped form or only in samples that are
in fact of the form $Ln$FePO$_{1-\delta}$, and to what degree sample quality effects the superconducting properties.

For LaFePO, values of \Tc\ that range from 3 K
\cite{kamihara_2006_1} to 7 K \cite{tegel_2008_1} have been
reported. Under pressure, we found that the \Tc\ of LaFePO reaches nearly 14 K at $\sim 110$ kbar~\cite{hamlin_2008_3}.  However, in a recent study of polycrystalline materials,
it was concluded that stoichiometric LaFePO is metallic but
non-superconducting \cite{mcqueen_2008_1} at temperatures as low
as 0.35 K.  Our recent work on single crystalline LaFePO
\cite{hamlin_2008_3} indicated \Tc\ values near 6 K as measured by
electrical resistivity and magnetization, but the specific heat
exhibited no discontinuity at \Tc\ leading us to propose that the
superconductivity may be associated with oxygen vacancies that
alter the carrier concentration in a small fraction of the sample,
although superconductivity characterized by an unusually small gap
value could not be ruled-out.  Yet, the superconductivity appeared
to be associated with the ZrCuSiAs structure, since the upper
critical field was anisotropic.  However, in a recent study of
single crystals of LaFePO \cite{fisher_2008_3}, a specific heat
anomaly roughly 60\% of the expected BCS value in magnitude at
the $T_c\approx 6$ K was reported, indicating that the
superconductivity can be a bulk phenomenon in this material.  These
samples were grown using a slightly different method than the one
we have employed, where La$_2$O$_3$ was used as the oxygen containing
precursor rather than Fe$_2$O$_3$, as used in our growths.  The
variation in the reported data indicate that the \Tc\ value and
superconducting fraction are sensitive to the details of the
synthesis conditions.

In this paper, we report the results of electrical resistivity,
magnetic susceptibility, and specific heat measurements on single
crystal samples of LaFePO, PrFePO, and NdFePO.  All three
compounds exhibit complete resistive superconducting transitions;
i.e., the resistivity drops  to a negligibly small value at low
temperatures.  However, as-grown crystals of PrFePO and NdFePO do
not show superconducting transitions in the magnetic
susceptibility and magnetic screening only develops after
post-annealing the samples in flowing O$_2$ at 700 $^{\circ}$C.
Annealing in flowing O$_2$ at 700 $^{\circ}$C for 24 hours is also
found to improve the superconductive properties of LaFePO.

\section{Experimental Details}
Single crystals of LaFePO, PrFePO, and NdFePO were grown from
elements and elemental oxides with purities $>99.9$\% in a molten
Sn:P flux. The growths took place over a 1 week period in quartz
ampoules which were sealed with 75 torr Ar at room temperature.
The inner surface of each quartz ampoule was coated with carbon by
a typical pyrolysis method. The starting materials were $Ln$, Fe$_2$O$_3$, P, and Sn, which were combined in the molar
ratios 9:3:6:80.5, similar to previous
reports~\cite{hamlin_2008_3,geibel_2008_1} for P-based
oxypnictide single crystals. The Fe$_2$O$_3$ powder was dried for
$\sim 10$ hours at 300 $^{\circ}$C before weighing. The ampoule
was heated to 1135 $^{\circ}$C at a rate of 35 $^{\circ}$C/hr,
kept at this temperature for 96 hours, and then rapidly cooled to
700 $^{\circ}$C. After removing the majority of the flux by
spinning the ampoules in a centrifuge, LaFePO, PrFePO, and NdFePO
single crystal platelets of an isometric form with typical
dimensions of $\sim0.5 \times 0.5 \times 0.05$ mm$^3$ or smaller
(particularly for the NdFePO crystals), were collected and cleaned
in hydrochloric acid to remove the flux from the surface of the
crystals prior to measurements. As previously reported~\cite{hamlin_2008_3}, the
platelets cleaved easily in the $ab$-plane and were notably
malleable, in contrast to the cuprate
superconductors. In order to explore the
effects of annealing in oxygen, several batches of crystals were
heated to 700 $^{\circ}$C for 24 hours under flowing O$_2$. Hereafter,
these samples will be referred to as ``O$_2$-annealed''.

X-ray powder diffraction measurements were made using a
diffractometer with a non-monochromated Cu K$\alpha$ source to
check the purity and crystal structure of the LaFePO and PrFePO
single crystals. Due to a small batch yield, x-ray diffraction
measurements were not made for the NdFePO crystals. As reported
previously, the crystals were difficult to grind into a fine
powder as a result of their malleability. Thus, the powder
diffraction patterns for $Ln$ = La and Pr crystals were generated
from a collection of several crystals which were cut into small
pieces using a razor blade and then ground into a coarse powder
using a mortar and pestle. The XRD results revealed diffraction
patterns that are consistent with those reported
previously~\cite{zimmer_1995_1}. Chemical analysis measurements
were also made using an FEI Company Model 600 scanning electron
microscope which revealed that the stoichiometry for the crystals
was consistent with the ratio 1:1:1:1 for $Ln$:Fe:P:O where $Ln$ = La, Pr, Nd.

Electrical resistivity $\rho(T)$ measurements were performed in a
four-wire configuration with the current in the $ab$-plane, at
temperatures $T$ $=$ 1.1-300 K using a conventional $^4$He
cryostat and a Quantum Design Physical Properties Measurement
System (PPMS). Due to the small size of the crystal samples and
uncertainties in the placement of the leads, the absolute
resistivity values are accurate only to $\sim \pm 50 \%$. Specific
heat $C(T)$ measurements were made for 2-300 K for LaFePO and
PrFePO crystals in a Quantum Design PPMS semiadiabatic calorimeter
using a heat-pulse technique. The specimens for specific heat
measurements were attached to a sapphire platform with a small
amount of Apiezon N grease and were composed of several hundred
single crystals for which the mass m = 13.06 mg (La unannealed),
8.93 mg (Pr unannealed), and 6.32 mg (Pr O$_2$-annealed). DC
magnetization $M(T,H)$ measurements were made using a Quantum
Design Magnetic Properties Measurement System (MPMS) in order to
probe both the superconducting and normal state properties of the
single crystal platelets. For the normal state measurements to
room temperature, the LaFePO and PrFePO specimens studied were the
same as those used for specific heat measurements and were mounted
in cotton packed gelatin capsules. For NdFePO, the normal state
measurements were made using a single crystal ($m$ = 0.0125 mg)
which was mounted on a piece of tape with the $ab$-plane
perpendicular to the applied magnetic field. To study the
superconducting state, individual single crystal specimens were
mounted with the $ab$-plane perpendicular to the magnetic field.
Multiple single crystal platelets were each individually measured
for 2-10 K and $H=5$ Oe under both zero field cooled (ZFC) and
field cooled (FC) conditions in order to characterize batch
homogeneity via variation in $T_c$, which was found to be minimal.

\section{Results}
\begin{figure}
    \begin{flushright}
    \includegraphics[width=13cm]{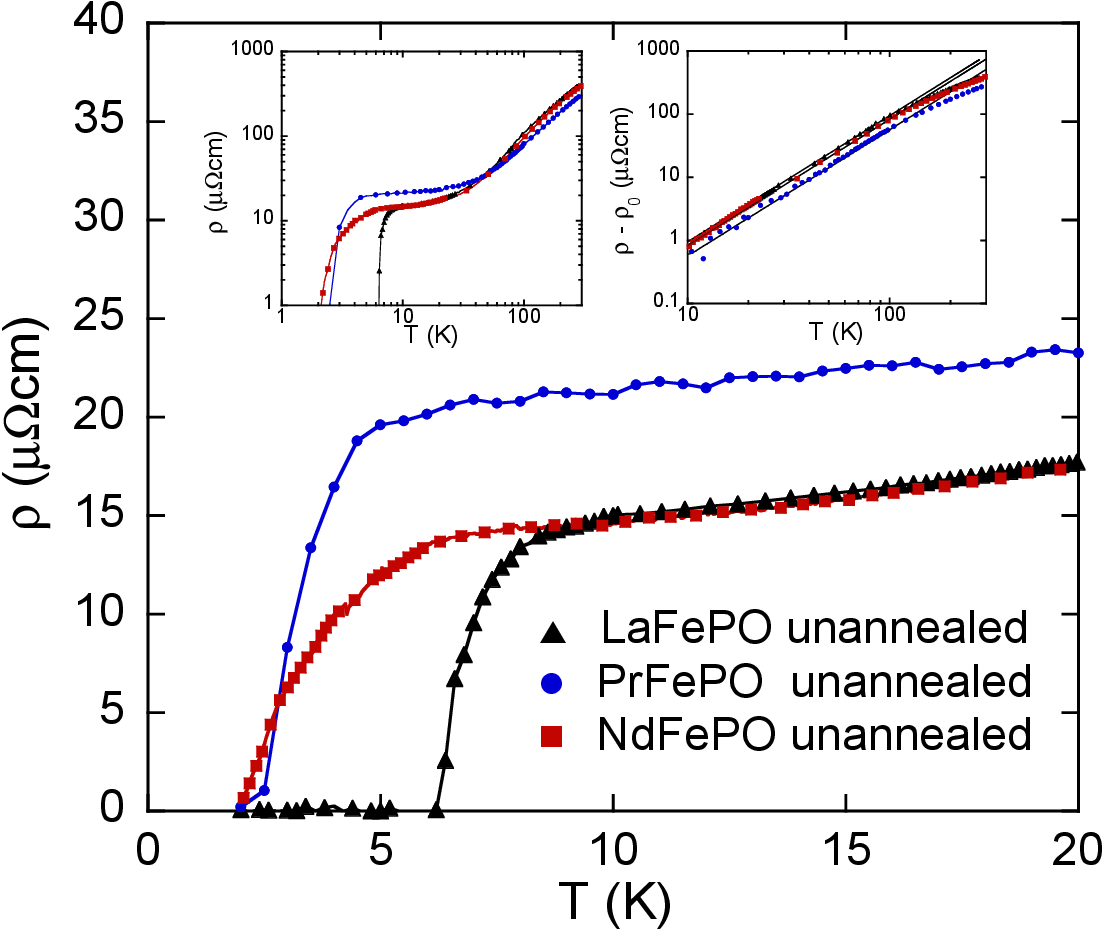}
    \end{flushright}
    \caption{Electrical resistivity $\rho$ versus temperature $T$ measured in the $ab$-plane for unannealed single crystals of LaFePO, PrFePO, and NdFePO. Left inset: Log-log plot of $\rho$ versus $T$ over the entire measured temperature range. Right inset: Log-log plot of $\rho - \rho _0$ versus $T$.  The solid lines are fits to the data which demonstrate the $T^2$ behavior for 10-100 K.}
        \label{fig:rho}
\end{figure}

The electrical resistivity data for unannealed $Ln$FePO ($Ln$ =
La, Pr, and Nd), shown in figure \ref{fig:rho}, reveal metallic
behavior where $\rho(T)$ decreases with decreasing $T$ until the
onset of the zero resistance state at the superconducting
transition temperatures $T_c \sim 6.6$ K, 3.2 K, and 3.1 K for
$Ln$ = La, Pr, and Nd respectively. The transition temperatures
are defined as the $T$ where $\rho(T)$ drops to 50\% of the
extrapolated normal state value. The transition widths are taken
as the difference in the temperatures where $\rho(T)$ drops to
10\% and 90\% of the extrapolated normal state value. For 100-300
K, the $\rho$($T$) data have approximately linear $T$ dependences
which evolve into quadratic forms for $T$ $\sim$ 10-100 K, as
shown in the right inset to figure \ref{fig:rho}.  Fits over this
temperature range show that the data are well described by the
expression,
\begin{equation}
 \label{eq:R}
  \rho(T) = \rho_0 + AT^2
\end{equation}
Additionally, the residual resistivity ratios $RRR=\rho(300
K)/\rho(0)$ for unannealed samples are large, reflecting the high
quality of the crystals. The results from fits to the $\rho(T)$
data are given in table \ref{table:rho}.

\begin{table*}
    \caption{A summary of results from electrical resistivity measurements for the compounds
    $Ln$FePO where $Ln$ = La, Pr, and Nd. The listed quantities are defined in the text and are
    abbreviated as the following: superconducting transition temperature $T_c$, width of the
    superconducting transition $\Delta T_c$(K), residual resistivity $\rho_0$ (equation \ref{eq:R}), quadratic coefficient $A$ (equation \ref{eq:R}),
    and residual resistivity ratio $RRR$ = $\rho_{300K}/\rho_0$.}
    \label{table:rho}
    \begin{flushright}
    \begin{tabular}{|c|c|c|c|c|c|}
    \hline                          $Ln$       & $T_c$(K)   & $\Delta T_c$(K)  & $\rho_0$($\mu \Omega$cm)  &$A$($\mu \Omega$cmK$^{-2}$)$\times 10^{-3}$ & $RRR$       \\
    \hline                          La        & 6.6        & 1.3             & 14.0                    & 9.46                                     & 32     \\
    \hline                          Pr        & 3.2        & 2.0             & 21.0                    & 5.66                                     & 14     \\
    \hline                          Nd        & 3.1        & 3.1             & 13.9                    & 8.28                                     & 28     \\
    \hline
    \end{tabular}
    \end{flushright}
\end{table*}

\begin{figure}
    \begin{flushright}
        \includegraphics[width=13cm]{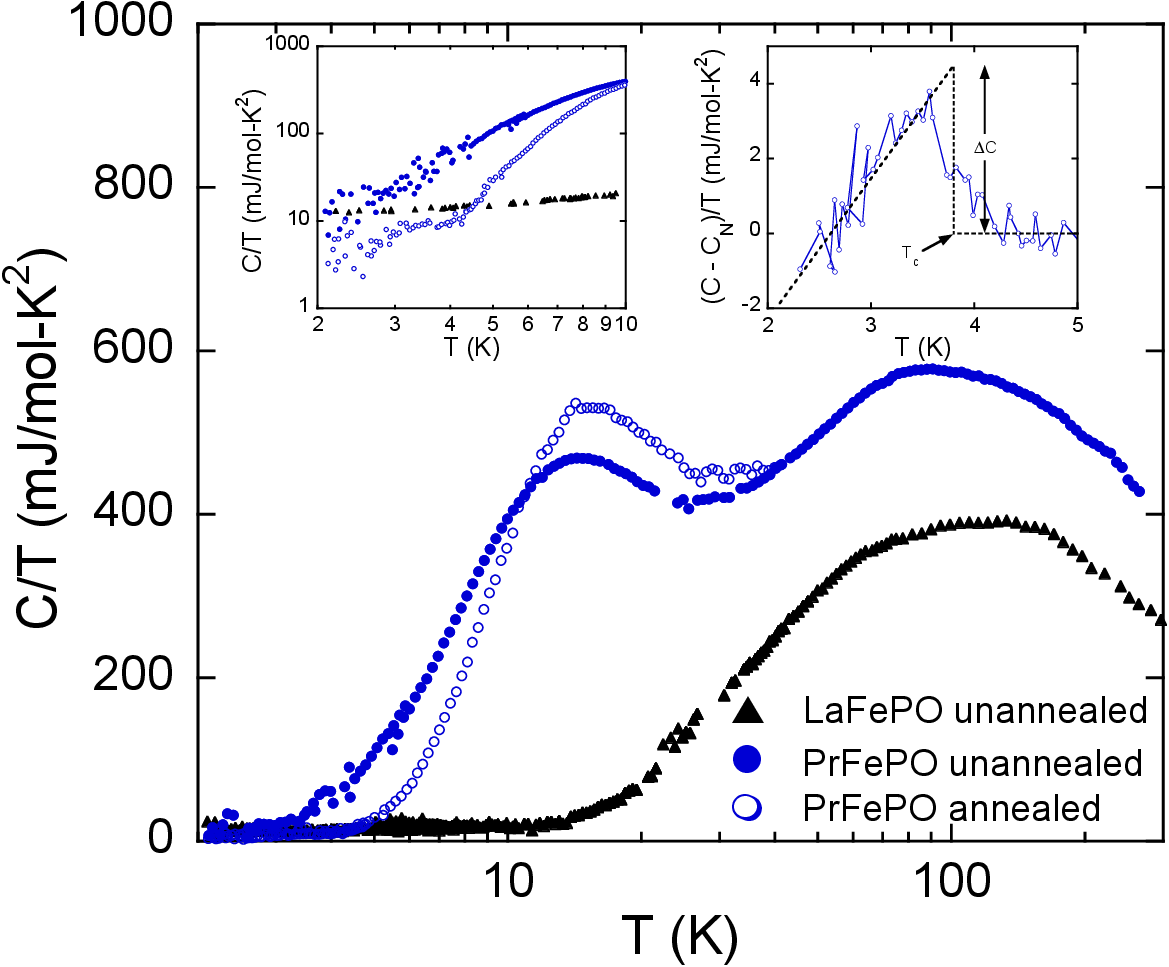}
    \end{flushright}
    \caption{Linear-log plot of specific heat divided by temperature $C/T$ versus
    temperature $T$ for unannealed LaFePO, unannealed PrFePO and O$_2$-annealed PrFePO.
    Left inset: Log-log $C/T$ versus $T$ plot for unannealed LaFePO, unannealed PrFePO and
    O$_2$-annealed PrFePO at low $T$. For O$_2$-annealed PrFePO, a broad jump is seen near
    $T_c$ = 3.6 K. Right inset: $C/T$ versus $T$ plot where the normal state electronic contribution
    has been subtracted, revealing the jump associated with the superconducting transition.
    The dotted line represents an equal entropy construction that was used to determine the value
    of $T_c$ and the size of the jump $\Delta C$.}
    \label{fig:C}
\end{figure}

Specific heat divided by temperature $C/T$ versus $T$ data for
unannealed LaFePO and PrFePO and O$_2$-annealed PrFePO are shown
in figure \ref{fig:C}. From room $T$, the unannealed $Ln$FePO
($Ln$ = La, Pr) specimens exhibit weak increases in $C/T$ with
decreasing $T$ that are followed by broad maxima near 110 K (La)
and 90 K (Pr). Below its maximum, $C/T$ for LaFePO continues to
decrease with decreasing $T$ and is described for 2 K $<$ $T$ $<$
8 K by a sum of electronic and lattice terms,
\begin{equation}
 \label{eq:Cp_T}
 C/T=\gamma+\beta T^{2}
\end{equation}
yielding an electronic specific heat coefficient $\gamma=12.7$
mJ/mol-K$^2$ and $\beta = r1944(T^3/\Theta_D)^3$J/mol-K, which
gives a Debye temperature $\Theta_D=268$ K.

\begin{figure}
    \begin{flushright}
        \includegraphics[width=13cm]{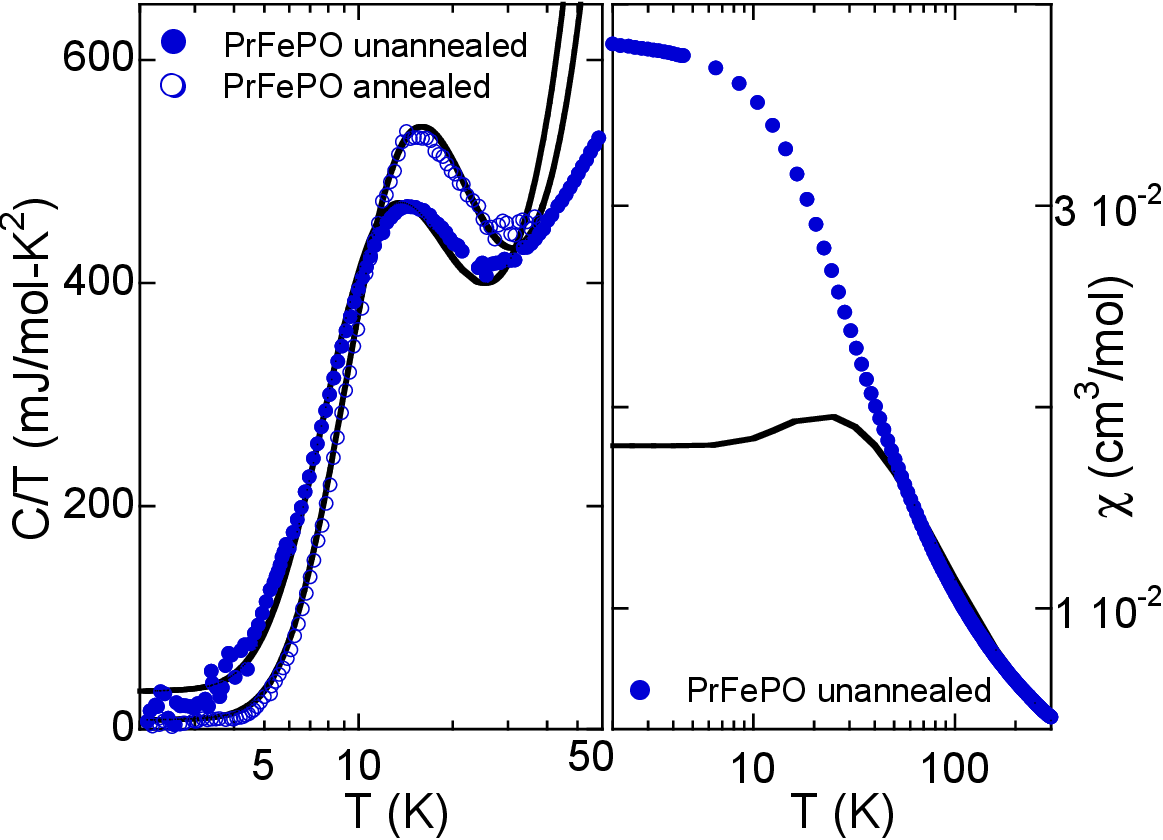}
    \end{flushright}
    \caption{Left panel: Specific heat divided by temperature $C/T$ versus $T$ for unannealed and O$_2$-annealed PrFePO. The solid lines are fits to the data using equation \ref{eq:Cp_T2}. Right panel: Magnetic susceptibility $\chi(T)$ = $M/H$ data for unannealed PrFePO. The solid line is a fit to the data using equation \ref{eq:sus}.}
    \label{fig:CEF}
\end{figure}

In contrast to results for LaFePO, $C/T$ versus $T$ data for
PrFePO specimens exhibit a trough-peak structure with minima near
27 K and 32 K and maxima near 14.5 K and 15.5 K for unannealed and
O$_2$-annealed samples, respectively. This type of behavior is
indicative of crystalline electric field (CEF) splitting of
energy levels of the Pr$^{3+}$ ions, which appears in specific
heat data as a so-called ``Schottky anomaly". Thus, for 2 K $<$
$T$ $<$ 45 K, the $C(T)/T$ data for both unannealed and
O$_2$-annealed PrFePO specimens were fit by an expression that
includes electronic, lattice, and Schottky terms given by,
\begin{equation}
 \label{eq:Cp_T2}
 C/T=\gamma+\beta T^{2}+rC_{\rm Sch}/T
\end{equation}
where $\gamma$ is the electronic specific heat coefficient and
$\beta T^2$ represents the lattice contribution to the specific
heat. The term $C_{\rm Sch}(T)$ is the Schottky specific heat
anomaly for a two-level system arising from the energy difference
between the CEF ground state and the first excited state, scaled
by a factor $r$, and is given by,
\begin{equation}
 \label{eq:schottky}
 C_{\rm Sch}(T)=R\left(\frac{\delta}{T}\right)^{2}\frac{g_{0}}{g_{1}}\frac{\exp
 (\delta)/T}{[1+(g_{0}/g_{1})\exp(\delta/T)]^{2}},
 \end{equation}
where $\delta$ is the energy difference in units of K between the
two levels, and $g_{0}$ and $g_{1}$ are the degeneracies of the
ground state and first excited state~\cite{gopal_book_1}. The
crystal structure of PrFePO belongs to the $p4/nmm$ space group,
with Pr$^{3+}$ ions at the points of a tetragonal unit
cell~\cite{zimmer_1995_1}. The energy level scheme for a Pr ion
for the tetragonal symmetry CEF Hamiltonian contains 5 singlets
and 2 doublets~\cite{michalski_2000_1}. As shown in figure
\ref{fig:CEF}, the best fit to the data suggests that both the
ground state and first excited states are nonmagnetic singlets. It is
also clear that, for unannealed specimens, equation \ref{eq:Cp_T2}
overestimates the value of $\gamma$. In contrast, the $C(T)/T$ data
for O$_2$-annealed samples are fit extremely well for 2 K $<$ $T$
$<$ 45 K and $\gamma$ is only slightly overestimated. These
results suggest that for the unannealed sample, there are a range
of CEF energy level splittings which broaden the Schottky peak and
render equation \ref{eq:Cp_T2} too simple an expression to
adequately describe the data. It is possible that a range of CEF
energy splittings could arise as the result of crystalline
disorder which would introduce variation in the local charge
distribution around individual Pr ions. Thus, it seems likely that
O$_2$-annealing reduces the spread of CEF energy level splittings,
resulting in a sharper Schottky peak and a better CEF fit for
O$_2$-annealed samples. For the unannealed and O$_2$-annealed
PrFePO samples, the splitting between the ground state and the
first excited state is 41.2 K and 48.4 K, respectively. The
remaining fit parameters are summarized in table 2.

As shown in the left inset to figure \ref{fig:C}, $\gamma$
approaches 12.7 and 10 mJ/mol-K$^2$ for unannealed LaFePO and
PrFePO, respectively, and 3 mJ/mol-K$^2$ for O$_2$-annealed
PrFePO, as $T$ goes to zero. For both unannealed LaFePO and
PrFePO, there is no detectable jump in the specific heat near the
superconducting transition temperatures inferred from $\rho(T)$.
In contrast, the O$_2$-annealed PrFePO specimen shows a broad jump
at low $T$. In order to determine $T_c$ and the size of the
specific heat jump $\Delta C$ for O$_2$-annealed PrFePO, the
normal state terms were subtracted from $C/T$ near the
discontinuity, resulting in the curve shown in the right inset to
figure \ref{fig:C}. The parameters, $T_c$ and $\Delta C$ were then
determined by applying an equal entropy construction, as shown by
the dashed line in the right inset to figure \ref{fig:C}. In
agreement with $\rho(T)$ for unannealed samples, $T_c$ = 3.6 K.
Interestingly, the specific heat jump $\Delta C$ = 15.1 mJ/mol-K
is nearly 100 \% of the value expected for a weak-coupling
conventional BCS superconductor, given by $\Delta C=1.52 \gamma
T_c$ = 16.4 mJ/mol-K if $\gamma$ is the extrapolated value of 3
mJ/mol-K$^2$ and $T_c$ = 3.6 K \cite{bardeen_1957_1}. In
comparison, unannealed LaFePO and PrFePO are expected to show
$\Delta C=127.4$ and 54.7 mJ/mol-K, respectively. By comparing the
expected specific heat jumps with the scatter in the $C(T)$ data
for the unannealed samples, it appears that at most $\sim 40 \%$
of the bulk is superconducting for the unannealed samples.

\begin{table*}\label{Ctable}
\caption{A summary of results from specific heat measurements for
the compounds $Ln$FePO where $Ln$ = La, Pr. For $\gamma$, both the
extrapolated $T=0$K value and the results from CEF fits (equation
\ref{eq:Cp_T2}) are reported. The additional parameters from fits
to the data using equation \ref{eq:Cp_T2} are the coefficient
$\beta$ of the phonon contribution, the splitting of the ground and first excited states
by the crystalline electric field $\delta$(K), and the
fitting parameter $r$.}
\begin{flushright}
\begin{tabular}{|c|c|c|c|c|c|}
\hline                          $Ln$               & $\gamma$(mJ/mol-K$^2$) & $\gamma_{CEF}$(mJ/mol-K$^2$)  & $\beta$(mJ/mol-K$^4$) & $\delta$(K)    & $r$ \\
\hline                          La unannealed               & 12.7                    &---& 0.10                 & ---            & ---  \\
\hline                          Pr unannealed     & 10     &33               & 0.28                 & 41.2           & 1.57 \\
\hline                          Pr O$_2$-annealed       & 3 & 7.3                    & 0.22                 & 48.4           & 2.27 \\
\hline
\end{tabular}
\end{flushright}
\end{table*}

\begin{figure}
    \begin{flushright}
        \includegraphics[width=13cm]{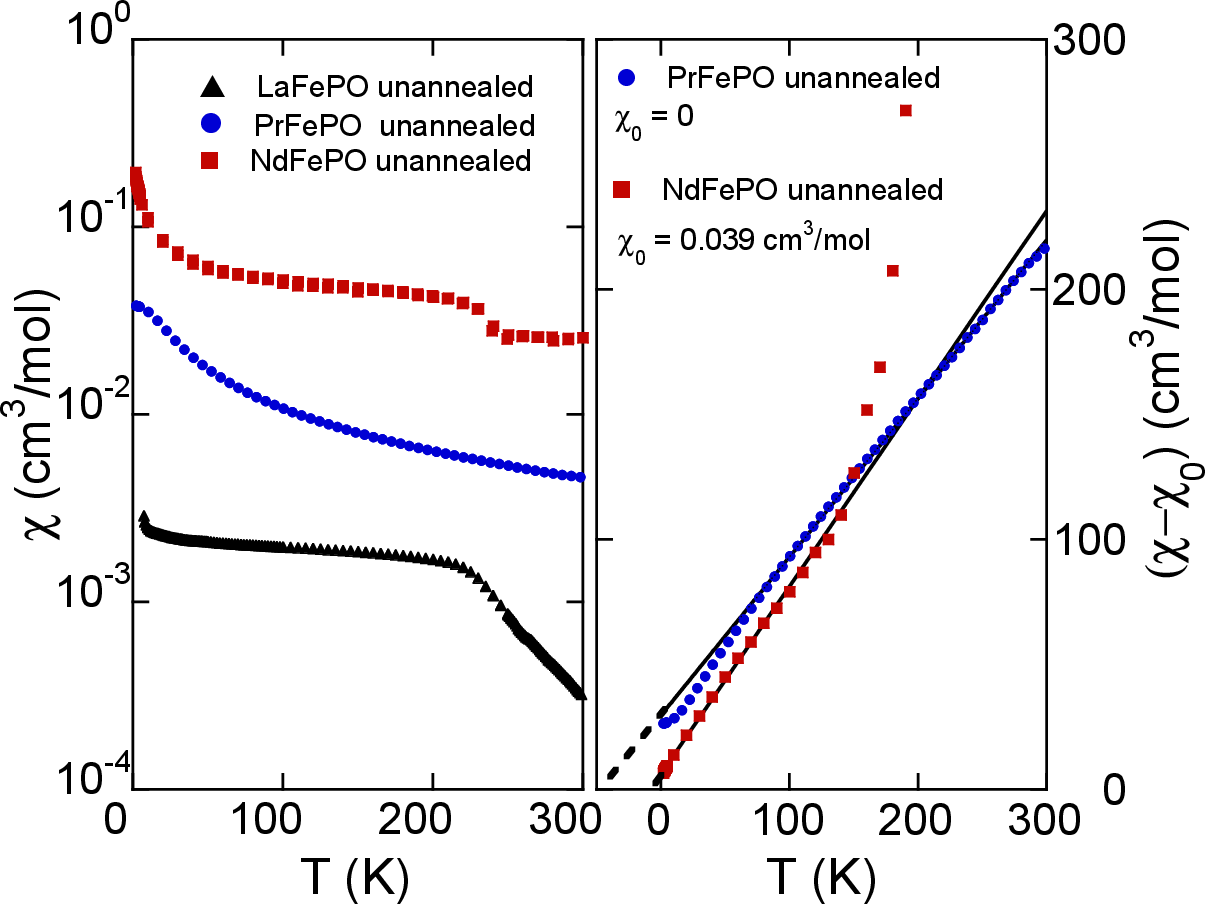}
    \end{flushright}
    \caption{Left panel: Magnetic susceptibility $\chi(T)$ = $M/H$ versus $T$ for $Ln$FePO ($Ln$ = La, Pr, Nd) data.
    Right panel: Inverse magnetic susceptibility $\chi^{-1}$ versus $T$ for $Ln$FePO ($Ln$ = Pr, Nd).
    The straight lines are fits to the data using modified Curie-Weiss expressions, as given by equation
    \ref{eq:CW}.}
    \label{fig:chi}
\end{figure}

The normal state $\chi(T)$ data for unannealed LaFePO, PrFePO, and
NdFePO are shown in figure \ref{fig:chi}. Near $\sim$ 220 K,
$\chi(T)$ increases strongly with decreasing $T$ for LaFePO and
NdFePO. This temperature dependence is similar to that of the
compound Fe$_2$P \cite{wold_1970_1}, which may be present as
inclusions or surface impurities.  We estimate that our observed
magnetic susceptibility is consistent with 1-2\% Fe$_2$P impurity
for both LaFePO and NdFePO. Interestingly, the $\chi$($T$) data
for PrFePO show no such feature. With decreasing $T$, $\chi(T)$
for LaFePO increases gradually and exhibits a low $T$ upturn which
persists down to 2 K. This behavior is not typical for Fe$_2$P,
and is either intrinsic to LaFePO or due to a small concentration
of some other paramagnetic impurity. For unannealed NdFePO,
$\chi$($T$) is consistent with Curie-Weiss behavior for Nd$^{3+}$
ions for 2 K $<$ $T$ $<$ 150 K, where the magnetic contribution
from the magnetic ordering at 220 K appears to be constant. For
this $T$ range, $\chi(T)$ can be described by a sum of a constant
term and Curie-Weiss function and is given by the expression,
\begin{equation}
 \label{eq:CW}
 \chi(T) = \chi_0 + C/(T-\Theta)
 \end{equation}
where $\chi_0$ = 0.039 cm$^3$/mol, $C$ = 1.33 cm$^3$K/mol, and
$\Theta$ = -8.1 K. The $\chi_0$ term is unusually large and it
seems likely it may be attributed to the magnetic contribution of
the impurity phase. From the Curie constant $C$, it appears that
the effective magnetic moment for the Nd ions is 3.26 $\mu_B$,
which is reasonably close to the value expected for Nd$^{3+}$ ions
according to Hund's rules ( $\mu_{eff}$ = 3.62 $\mu_B$ for Nd$^{3+}$).
Lastly, the negative Curie-Weiss temperature indicates
antiferromagnetic spin correlations for the Nd ions.

Similarly, $\chi(T)$ for unannealed PrFePO shows behavior that is
consistent with Curie-Weiss behavior for 50 K $<$ $T$ $<$ 300 K
and is described by equation \ref{eq:CW}, where $\chi_0$ = 0, $C$
= 1.58 cm$^3$K/mol, and $\Theta$ = -46.7 K. Again, the Curie
constant yields a reasonable value of $\mu_{eff}$ = 3.55 $\mu_B$
($\mu_{eff}$ = 3.58 $\mu_B$ for Pr$^{3+}$ according to Hund's rules),
while the large negative Curie-Weiss temperature indicates
antiferromagnetic spin correlations for the Pr ions. For $T$ $<$
50 K, $\chi$($T$) is suppressed from Curie-Weiss behavior and
saturates to a constant value with decreasing $T$. This result is
consistent with the ground and first excited states being
nonmagnetic singlets, between which the energy splitting is close
to 41 K as inferred from the fit to the Schottky anomaly in the
$C(T)/T$ data. In order to describe the effect of crystalline
electric field splitting on $\chi(T)$, a CEF fitting scheme is
considered. As mentioned above, the crystal structure of PrFePO
belongs to the $p4/nmm$ space group, with Pr$^{3+}$ ions at the
points of a tetragonal unit cell.  The crystalline electric field
(CEF) Hamiltonian of the Pr ions has the form
\begin{equation}
 \label{eq:CEF}
 {\cal H}_{\rm CEF}=B_{2}^{0}O_{2}^{0}+B_{4}^{0}O_{4}^{0}+5B_{4}^{0}O_{4}^{4}+B_{6}^{0}O_{6}^{0}-21B_{6}^{0}O_{6}^{4},
\end{equation}
where $O_{2}^{0}$, $O_{4}^{0}$, $O_{4}^{4}$, $O_{6}^{0}$, and
$O_{6}^{4}$ are Stevens operators for the $J$ = 4 manifold and the
$B$s are parameters determined from experiment.  The CEF
Hamiltonian ${\cal H}_{\rm CEF}$ splits the $J$ = 4 Hund's rule
multiplet of Pr$^{3+}$ into 5 singlets and 2 doublets.  In the
presence of an external magnetic field, the Zeeman interaction
mixes and splits the CEF energy levels and the magnetic
susceptibility is given by~\cite{Vleck32}
\begin{equation}\label{eq:sus} \chi_{\rm
CEF}=\frac{\sum_{n}\left[(E_{n}^{(1)})^{2}/kT-2E_{n}^{(2)}\right]\exp\left(-E_{n}^{(0)}/kT\right)}{\sum_{n}\exp\left(-E_{n}^{(0)}/kT\right)},
\end{equation}
where the $E_{n}^{(0)}$ are the unperturbed cubic CEF levels,
$E_{n}^{(1)}=\mu_{B}g\langle\phi_{n}|J|\phi_{n}\rangle$ with
$\phi$ the CEF wave functions and $g$ the Land{\'e} g-factor, and
\begin{equation}
E_{n}^{(2)}=\sum_{n^{'}\neq
n}\mu_{B}^{2}g^{2}\frac{|\langle\phi_{n}|J|\phi_{n}^{'}\rangle|^{2}}{E_{n}^{(0)}-E_{n^{'}}^{(0)}}.
\end{equation}
In the molecular-field approximation, the measured magnetic
susceptibility $\chi$ is given by $\chi=\chi_{\rm CEF}/(1-\lambda
\chi_{\rm CEF})$, where $\lambda$ is the molecular field parameter
that describes exchange interactions between Pr$^{3+}$
ions~\cite{Murao57}. Our best fit is shown as the solid curve in
figure \ref{fig:CEF} where it is apparent that the CEF fit only
describes the data qualitatively. The CEF fit parameters are
$B_{2}^{0}$ = -1.1 meV, $B_{4}^{0}$=0.0001 meV, $B_{4}^{4}$ = 0.0082
meV, $B_{6}^{0}$ = -0.0030 meV, $B_{6}^{4}$ = -0.0148 meV, and
$\lambda$ = -18.4 mol/emu.  This set of CEF parameters suggests the
energy separation between the ground and first excited states is
44.3 K, and both ground and first excited states are singlets, in
good agreement with the specific heat data.

\begin{figure}
    \begin{flushright}
        \includegraphics[width=13cm]{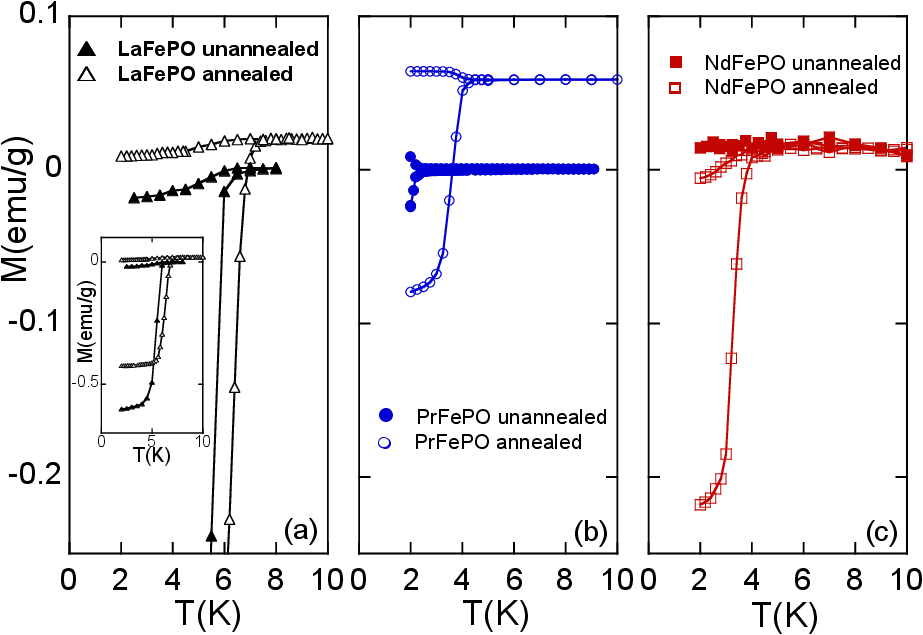}
    \end{flushright}
    \caption{Left panel: Magnetization $M(T)$ data for $H$ = 5 Oe for unannealed and O$_2$-annealed LaFePO samples.
    Center panel: Magnetization $M(T)$ data for $H$ = 5 Oe for unannealed and O$_2$-annealed PrFePO samples.
    Right panel: Magnetization $M(T)$ data for $H$ = 5 Oe for unannealed and O$_2$-annealed NdFePO samples.}
    \label{fig:chi_lowT}
\end{figure}

\begin{figure}
    \begin{flushright}
        \includegraphics[width=13cm]{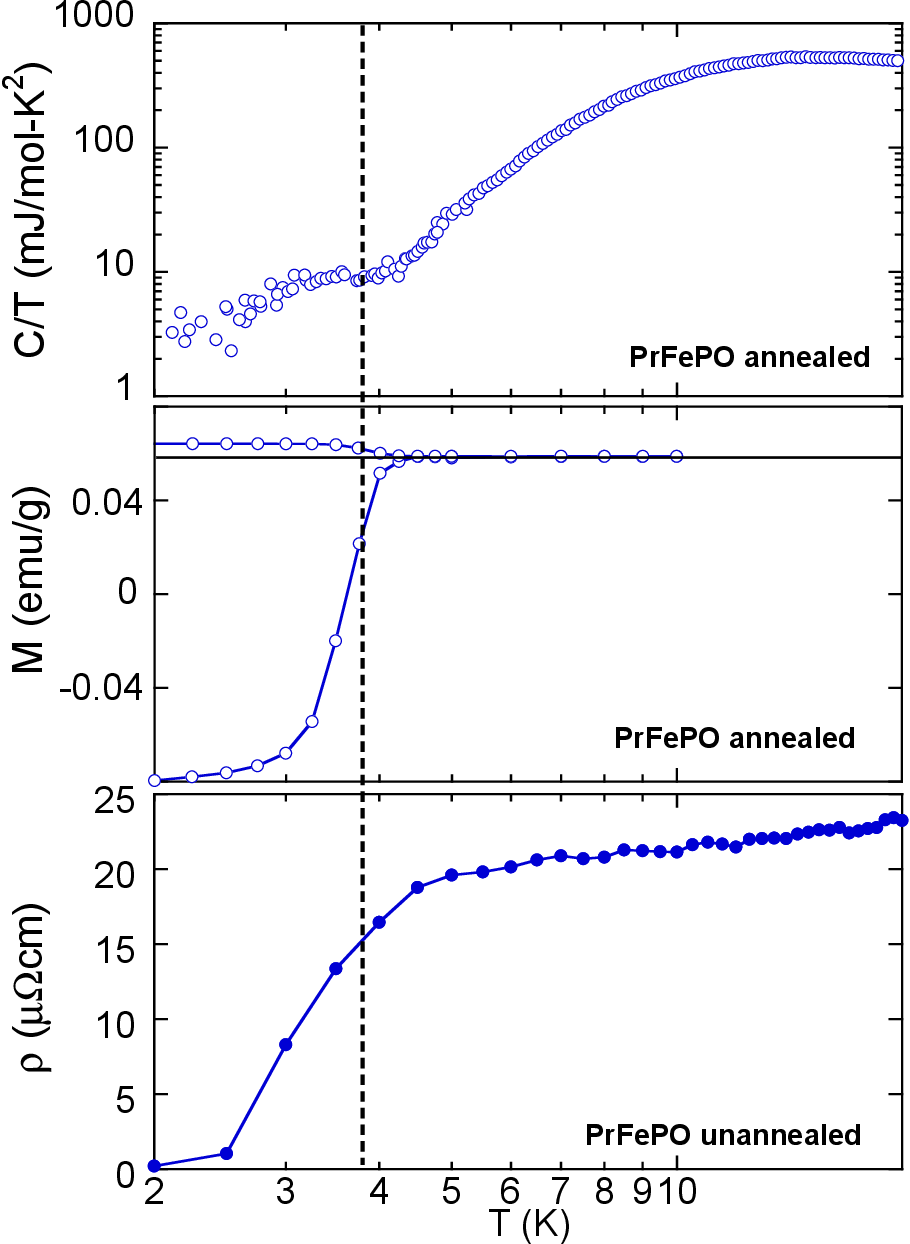}
    \end{flushright}
    \caption{A comparison of the electrical resistivity $\rho$ versus $T$ data for unannealed PrFePO samples with
    specific heat divided temperature $C/T$ versus $T$ and magnetization $M$ versus $T$ for O$_2$-annealed PrFePO showing the agreement in $T_c$ among the measurements.}
    \label{fig:comparison}
\end{figure}

In an effort to determine whether the superconductivity is a bulk
phenomenon, zero-field-cooled (ZFC) and field-cooled (FC)
measurements of the magnetic susceptibility were made in a field
of 5 Oe for unannealed and O$_2$-annealed LaFePO, PrFePO, and
NdFePO crystals. It should be noted at the outset that the values
of $M$(emu/g) are accurate to roughly $\pm 15\%$ due to
uncertainties in mass and geometry of the
small crystals. A plot of the ZFC and FC magnetic susceptibility through the superconducting
transition temperatures observed in $\rho$($T$) is shown in
figure \ref{fig:chi_lowT}. For unannealed LaFePO, the onset
temperature for LaFePO is near 6.0 K. With O$_2$-annealing, $T_c$
increases to a value near 7.0 K. For both the unannealed and
O$_2$-annealed cases, the FC data return to $\sim$ 5 \% of the ZFC
values. It also appears that the normal state magnetization is
slightly enhanced for the O$_2$-annealed samples. Unannealed
PrFePO shows a weak superconducting transition near 2 K for ZFC
measurements and the FC measurement reveals a weak upturn in the
magnetization below $T_c$. Although this feature is anomalous for
typical superconductors, it has been repeatedly observed for these
crystals. With annealing, PrFePO develops a pronounced
superconducting transition with an onset near $T_c$ = 4.4 K which
is again accompanied by the anomalous low $T$ upturn in the FC
measurement. Interestingly, the normal state magnetization is
markedly enhanced for the O$_2$-annealed samples. A comparison of our specific heat, magnetic susceptibility, and resistivity measurements on PrFePO is shown in figure~\ref{fig:comparison}.  Unannealed
NdFePO shows no evidence for a superconducting transition down to
2 K. Again, O$_2$-annealing promotes superconductivity, yielding
an onset temperature $T_c$ = 4.4 K. For O$_2$-annealed NdFePO, the
FC data return to $\sim$ 3 \% of the ZFC values. Also, there is no
offset in the normal state magnetization as the result of
O$_2$-annealing. Thus, it is evident that O$_2$-annealing acts to
promote the superconducting state for LaFePO, PrFePO, and NdFePO,
although it is unclear what the overall effect of O$_2$-annealing
is on the normal state behavior. Additionally, the small recovery
of the diamagnetic signal on field-cooling indicates either that
the material shows strong vortex pinning or that the
superconductivity is not a bulk phenomenon. Further studies are
required to make this distinction.

\section{Discussion}
\begin{figure}
    \begin{flushright}
        \includegraphics[width=13cm]{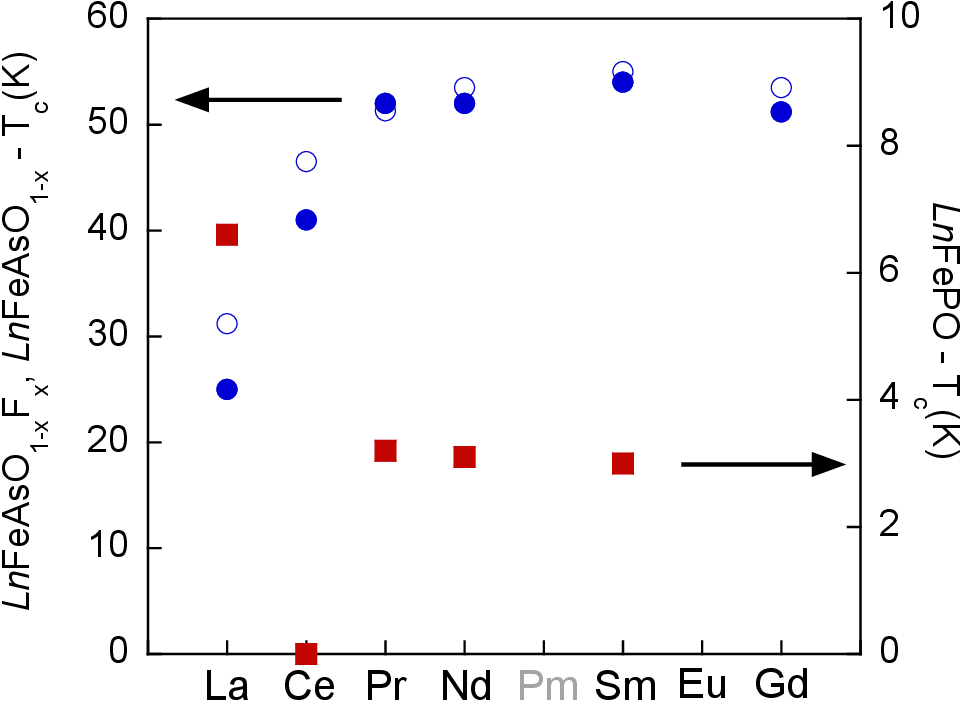}
    \end{flushright}
    \caption{A comparison of the evolution of the superconducting transition temperature $T_c$ versus lanthanide $Ln$ for the series $Ln$FePO with $Ln$ = La, Ce \cite{bruning_2008_1}, Pr, Nd, Sm \cite{kamihara_2008_2} (solid squares), the optimally fluorine doped compounds $Ln$FeAsO$_{1-x}$F$_x$ with $Ln$ = La \cite{kamihara_2008_1}, Ce \cite{chen_2008_1}, Pr \cite{ren_2008_3}, Nd \cite{ren_2008_2}, Sm \cite{ren_2008_1}, and Gd \cite{yang_2008_1} (solid circles), and the oxygen deficient compounds $Ln$FeAsO$_{1-\delta}$ \cite{yang_2008_1} (open circles).}
    \label{fig:compareTc}
\end{figure}

An important question for understanding the Fe-$Pn$ superconductors is how the magnetism associated with the $Ln$ ion influences the superconductivity. For conventional superconductors, it is expected that \Tc\ values are suppressed by the introduction of magnetic ions, as is observed for the well known case of de Gennes' scaling in BCS superconductors~\cite{degennes_1962_1,maple_1978_1}. Interestingly, both the series $Ln$FeAsO$_{1-x}$F$_x$ and $Ln$FeAsO$_{1-\delta}$ display an increase in the maximum \Tc\ values on replacing La with magnetic $Ln$ ions~\cite{kamihara_2008_1,chen_2008_1,ren_2008_3,ren_2008_2,ren_2008_1}. This result, together with the apparent suppression of a spin density wave that is correlated with the appearance of superconductivity, has led to the suggestion that the superconductivity exhibited by these compounds is unconventional in the sense that it is promoted by magnetic interactions \cite{drew_2008_1}. In contrast, for the series $Ln$FePO, the \Tc\ values decrease upon replacing La with Pr, Nd, or Sm, as could be expected for a conventional superconductor in the presence of magnetic pairbreaking interactions. For the purpose of comparison, the evolution of $T_c$ versus $Ln$ for the series $Ln$FeAsO$_{1-x}$F$_x$, $Ln$FeAsO$_{1-\delta}$, and $Ln$FePO is shown in figure~\ref{fig:compareTc}. While one should be careful when comparing systematics between undoped $Ln$FePO and optimally doped $Ln$FeAsO$_{1-x}$F$_x$ and $Ln$FeAsO$_{1-\delta}$ samples, the opposite effects of magnetic ion substitution could be taken as evidence for a difference between the P- and As-based compounds in the mechanism of superconductivity. The point of view that the superconductivity for the series $Ln$FePO is BCS-like is supported by the lack of anomalous behavior in the electrical resistivity, magnetic susceptibility, and specific heat.

Another important question for understanding the Fe-$Pn$
superconductors is how sample quality affects the
superconductivity. Broadly speaking, sample quality can be defined
as either crystalline or chemical homogeneity, each of which has
the potential to affect the physical properties in unique ways. It
has been noted \cite{kamihara_2008_3} that among several samples
of undoped LaFePO, those with the lowest residual resistivities
display the highest \Tc\ values.  A correlation between low
residual resistivity and high \Tc\ has also been observed in
SmFePO \cite{kamihara_2008_2}.  This trend could derive from
several possible differences among the samples, including lattice
disorder, oxygen vacancy induced changes in the carrier density, or oxygen vacancy induced changes in the lattice parameters. Our measurements for LaFePO, PrFePO, and NdFePO, which
clearly show that O$_2$-annealing promotes the superconducting
state, suggest a correlation between sample quality and higher \Tc\ values. This is shown in figure \ref{fig:comparison} for PrFePO where
electrical resistivity for unannealed samples, magnetization data
for O$_2$-annealed samples, and specific heat data for
O$_2$-annealed samples are compared. Comparison of magnetization
data for unannealed and O$_2$-annealed samples also show that
superconductivity is promoted by O$_2$-annealing (figure
\ref{fig:chi_lowT}). In general, annealing tends to improve
crystalline quality by allowing disorder in the lattice to relax,
in addition to reducing possible chemical inhomogeneity. On the
other hand, annealing can also promote phase separation. For this
reason, although it seems reasonable to correlate crystalline
quality with an enhancement of the superconducting state, it is
still necessary to carry out a more systematic study of the effect
of annealing in these compounds. Furthermore, we note that it may
be possible to improve the superconducting properties of the Fe-As
compounds by synthesizing high quality single crystal specimens.

\section{Concluding Remarks}
We have reported results for single crystals of the compounds LaFePO,
PrFePO and NdFePO, which were prepared by means of a flux growth
technique. Measurements of electrical resistivity, magnetic
susceptibility and specific heat reveal metallic behavior for
unannealed samples, which evolves into a superconducting state at
low $T$ in $Ln$ = La ($T_c$ = 6.6 K), $Ln$ = Pr ($T_c$ = 3.2 K),
and $Ln$ = Nd ($T_c$ = 3.1 K). The effect of annealing in flowing
O$_2$ at 700 $^{\circ}$C for 24 hours was also studied and found to enhance the superconducting properties. Although it seems likely that the positive effect of
O$_2$-annealing on the superconducting properties is due to
improved crystallinity and chemical homogeneity, further research
will be required to support this conclusion. Finally, the evolution of
$T_c$ with $Ln$ ion for the P- and As-based $Ln$Fe$Pn$O compounds
appears to be in opposite directions: \Tc\ increases as $Ln$ is
varied from La to Sm for $Pn$ = As and decreases as $Ln$ is varied
from La to Sm for $Pn$ = P.  This suggests that the mechanism of the superconductivity in these
two series of compounds may be different; i.e., the superconductivity may be unconventional for the arsenides and conventional for the phosphides.

\subsection{Acknowledgments}
The crystal growth work was supported by the U.~S.\ Department of Energy (DOE) under research Grant DE~FG02-04ER46105 and acquisition of crystal growth equipment was sponsored by the U.~S.\ DOE through Grant DE~FG02-04ER46178.  Low temperature measurements were funded by the National Science Foundation (NSF) under Grant 0802478.  The work of N.\ Crisosto was supported by the NSF through the Research Experience for Undergraduates (REU) program under grant PHY-0552402.  The authors thank C.~A.\ McElroy for experimental assistance with the specific heat measurements.

\end{document}